\documentclass[aps,prd,twocolumn,superscriptaddress,nofootinbib]{revtex4-1}

\usepackage{latexsym}
\usepackage{amsmath}
\usepackage{amssymb}
\usepackage{amsfonts}
\usepackage{bm}
\usepackage{physics}

\usepackage{color}
\definecolor{purple}{rgb}{0.5,0,0.5}
\definecolor{blue}{rgb}{0.0,0,0.9}
\definecolor{prdblue}{rgb}{0.133,0.118,0.498}
\usepackage[colorlinks=true, pdfstartview=FitV, linkcolor=prdblue, citecolor= prdblue, urlcolor=prdblue]{hyperref}

\usepackage{supertabular} 
\usepackage{placeins}
\usepackage{epsfig}
\usepackage{graphicx}
\usepackage{slashed}
\begin{document}

\title{Landau-Khalatnikov-Fradkin 
Transformations in Quantum Electrodynamics: \\  For Perturbation Theory and Dynamical Mass Generation}

\author{Anam Ashraf}
\email[]{anam.ashraf@au.edu.pk}
\affiliation{Department of Physics, Quaid-i-Azam University, Islamabad 45320, Pakistan.}
\affiliation{Department of Physics, Air University, Islamabad, Pakistan.}

\author{M. Jamil Aslam}
\email[]{jamil@qau.edu.pk}
\affiliation{Department of Physics, Quaid-i-Azam University, Islamabad 45320, Pakistan.}

\author{Faisal Akram}
\email[]{faisal.chep@pu.edu.pk}
\affiliation{Centre for High Energy Physics, University of the Punjab, Lahore (54590), Pakistan.}

\author{Adnan Bashir}
\email[]{adnan.bashir@dci.uhu.es;
adnan.bashir@umich.mx}
\affiliation{Dpto. Ciencias Integradas, Centro de Estudios Avanzados en Fis., Mat. y Comp., \\
Fac. Ciencias Experimentales, Universidad de Huelva, Huelva, 21071, Spain.}
\affiliation{Instituto de F\'isica y Matem\'aticas, Universidad
Michoacana de San Nicol\'as de Hidalgo, \\
Morelia, Michoac\'an
58040, M\'exico}

\date{\today}

\begin{abstract}

We carry out a comprehensive analysis of the Landau-Khalatnikov-Fradkin transformations for a charged fermion propagator at the two-loop level in quantum electrodynamics (QED). Starting with an arbitrary covariant gauge $\xi$ and space-time dimension $d$, we provide its explicit expressions in three and four-dimensional QED. We begin with the tree-level fermion propagator in the Landau gauge and  gauge-transform it to obtain an analytical expression for an all order result in an arbitrary covariant gauge. We expand it out to two-loops both for the massless and massive propagators in three and four space-time dimensions. In addition to comparing with all earlier results in the literature wherever possible, we also study constraints of multiplicative renormalizabilty of our results in four-dimensional QED which are logarithmically divergent. 
Finally, we analyze representative solutions of the fermion propagator which correspond to dynamical chiral symmetry breaking and mass generation in QED. We study the gauge dependence of these emergent solutions, that of the Euclidean pole mass and the chiral fermion condensate.

\end{abstract}


\maketitle


\section{Introduction}

All fundamental interactions which constitute the celebrated Standard Model of Salam, Weinberg and Glashow, and which orchestrate the dance of events in the microscopic world, are gauge field theories. Gauge symmetry of these theories leads to identities and transformations which are verifiable at every order of perturbation theory. As their derivation makes no reference to the strength of the interactions involved, we expect the validity of these identities and transformations to remain intact even if novel non-perturbative solutions emerge which remain inaccessible in perturbation theory. These emergent phenomena arise when interactions are strong enough to render perturbative expansion meaningless. Nonetheless, it is true that two of the three fundamental interactions remain parturbative at all energies accessible in current experiments. Moreover, the third interaction, quantum chromodynamics (QCD), is also perturbative at high energies and becomes asymptotically free. We  explore both the perturbative and (hypothetical) non-perturbative domains of quantum electrodynamics (QED) in this article.

The more well-known consequence of local gauge covariance in QED are the Ward-Fradkin-Green-Takahashi identities~(WFGTI)~\cite{Ward:1950xp, Fradkin:1955jr,Green:1953te,Takahashi:1957xn}. These identities relate various $n$-point Green functions to each other.  Although these are intrinsically non-perturbative in nature, they are satisfied at every order in perturbation theory and serve as a check of the computation in an arbitrary covariant gauge. As an example, the electron propagator is related to the electron-photon interaction vertex through a WFGTI. The generalization of these identities to QCD are the Slavnov-Taylor (ST) identities,~\cite{Slavnov:1972fg,Taylor:1971ff}. 

Another consequence of gauge covariance in QED are the transformations which allow us to relate a given Green function in one gauge to its value in another gauge. These relatively less known transformations for electron and photon propagators, as well as the electron-photon vertex in different covariant gauges were derived initially by Landau and Khalatnikov in Ref.~\cite{Landau:1955zz} and then independently by Fradkin in Ref.~\cite{Fradkin:1955jr} 
by employing canonical formalism of QED. Therefore, these are rightfully named as Landau-Khalatnikov-Fradkin (LKF) transformations. These transformations were originally written in position space. The general rules which govern these transformations, although visibly compact, are by no means simple. These transformations were later derived by Johnson and Zumino through functional methods~\cite{Johnson:1959zz,Zumino:1959wt}. Similarly to the ST identities, the QED transformations were generalized to the correlations functions of QCD a few  decades later in Refs.~\cite{Aslam:2015nia,DeMeerleer:2018txc,DeMeerleer:2019kmh} by Aslam {\em et. al.} and Dall'Olio {\em et. al.}, respectively. 
 Noticeably, despite the fact that the WFGTI and the LKF transformations are independent relations, yet it is established that if the WFGTI holds in one gauge, then if the Green functions involved are transformed to another gauge under LKF transformation, the WFGTI in that other gauge will be satisfied automatically. Due to the momentum space nature of WFGTI, these identities have been extensively implemented in the gauge studies in the literature.
However, LKF transformations have not had the same amount of applications. 

However, one can still find numerous applications of the gauge transformations in the literature in the context of  QED in different space-time dimensions: QED2~\cite{Nicasio:2023zec}, QED3~\cite{Burden:1993gy,Aitchison:1997ua,Bashir:1999bd,Bashir:2000rv,Bashir:2000ur,Bashir:2002sp,Bashir:2004yt,Bashir:2004rg,Bashir:2005wt,Bashir:2009fv,Kotikov:2023qdm,Gusynin:2020cra},
QED4 or QED~\cite{Curtis:1990zs,Bashir:1994az,Bashir:2002sp,Kizilersu:2009kg}, QED$d$~\cite{Bashir:2004hh,Jia:2016wyu}, scalar QED~\cite{Villanueva-Sandoval:2013opv,Ahmadiniaz:2015kfq,Fernandez-Rangel:2016zac,Kotikov:2019bqo} and reduced QED~\cite{Ahmad:2016dsb,James:2019ctc,Albino:2022efn}, in perturbation theory as well as in the non-perturbative domain of QED. Several of these works are based upon seeking analytical insight into the non-perturbative structure of the fermion-photon vertex by demanding
correct gauge covariance properties of the massless fermion propagator.
The remaining works study how a fermion propagator would transform under a variation of gauge, analyzing its  perturbative expansion. The results thus obtained for the LKF transformed fermion propagator can be verified wherever perturbative results are available for that order. On the other hand, a non-perturbative analysis provides us insight into the gauge invariance of chiral symmetry breaking, an important topic in its own right.

At the perturbative level, employing the LKF transformations, we calculate  fermion propagator to two loops both for the massless and massive cases in three and four space-time dimensions. Our starting point is the tree-level propagator in the Landau gauge, which remains a valid starting point even at the one-loop level,~\cite{Dong:1994jr}. 
However, at the the two-loop level, fermion propagator violates the so-called transversality condition~\cite{Dong:1994jr,Burden:1998gr}. We confirm that the gauge parameter dependent pieces which should be reproduced 
accurately do so in the cases where perturbative results are known. 
For the four-dimensional case of ultraviolet divergences, we also provide renormalized results. Note that all previous calculations within the framework of these transformations have been carried out up to one-loop order except massless QED3~\cite{Bashir:2000ur}.

 Although QED manifests a perturbative behavior at all
observable scales, it has been theoretically observed that a toy QED with an artificially scaled up coupling exhibits the phenomenon of dynamical mass generation. Such a phase transition from a massless to a massive electron has long been studied, see for example~\cite{Fomin:1983kyk,Miransky:1984ef,Miransky:1985gbh} for the first articles in the field. For a more recent recap, consult~\cite{Curtis:1993py,Bashir:2011dp,Kizilersu:2014ela,Albino:2018ncl}.
This phase transition is characterized by a critical coupling, $\alpha_c \simeq 1$, above which dynamical chiral symmetry breaking and mass generation for electrons takes place.
 the corresponding non-perturbative treatment of the LKF transformations, we provide a glimpse by assuming a representative solution which has the known infrared enhancement of the fermion mass function and falls-off as $1/p^2$ as observed in perturbation theory. 
We do not include any scaling violations in the input solution. We then numerically transform this solution to other gauges and analyze the gauge (in)dependence of the mass-function, fermion condensate and the Euclidean pole mass.

We have organized this study as follows. In Sect.~\ref{LKFEP}, we
provide the generalities of the formalism and outline the procedure we use throughout the manuscript. This section also sets the notation for all the subsequent sections.
In Sect.~\ref{MLMFP}, we present the results obtained directly from the LKF transformations till the two-loop order for the fermion propagator for QED3 and QED4, both for the massless and massive cases. This relatively longer section is followed by two shorter sections. In Sect.~\ref{renormalization}, we outline our renormalization procedure, and provide the multiplicatively renormalized results for the wavefunction renormalization and the mass function in QED4. This is followed by a cursory  exploration into the gauge dependence of the dynamically generated mass function, the chiral fermion condensate and the Euclidean mass pole in Sect.~\ref{DGM}. Finally, Sect.~\ref{conclusion} summarizes the main findings of this work and provides a glimpse into possible avenues of future research.

\section{The LKFT and fermion propagator} \label{LKFEP}

We begin by recalling the defining expressions for the LKF transformations written in the coordinate space, the fermion propagator both in the position and momentum spaces and the Fourier transforms we shall use all throughout this article. 

\subsection{Generalities of the formalism}

In a covariant gauge, characterized by the parameter $\xi$, the most general form of the Euclidean space fermion propagator in momentum space can be written in the following equivalent ways\,: 
\begin{equation}
S_{F}\left(p;\xi\right)=A\left(p;\xi\right)+\frac{iB\left(p;\xi\right)}{\slashed{p}}=\frac{F\left(p;\xi\right)}{i\slashed{p}-\mathcal{M}\left(p;\xi\right)}\,,\label{eq:1}
\end{equation}
where $F\left(p;\xi\right)$ and $\mathcal{M}\left(p;\xi\right)$ are called the
wavefunction renormalization and the mass function, respectively. The corresponding Lorentz decomposition in the coordinate space reads as\,:
\begin{equation}
S_{F}\left(x;\xi\right)=\slashed{x}X\left(x;\xi\right)+Y\left(x;\xi\right) \,.\label{eq:2}
\end{equation}
The position and momentum space functions are related to each other through the Fourier
transformation\,: 
\begin{eqnarray}
S_{F}\left(x;\xi\right)&=&\int\frac{d^{d}p}{\left(2\pi\right)^{d}}\text{e}^{-ip\cdot x}S_{F}\left(p;\xi\right)\,,\label{eq:3a}\\
S_{F}\left(p;\xi\right)&=&\int d^{d}x\text{e}^{ip\cdot x}S_{F}\left(x;\xi\right)\,,\label{eq:3b}
\end{eqnarray}
where $d$ denotes the dimension of space-time, and $p\cdot x$ represents the corresponding Euclidean scalar product. Taking  $\theta$ to be the angle between $p$ and $x$, and integrating over the remaining $(d-2)$ angles, we can write the measure $d^{d}p = p^{d-1}dp\sin^{d-2}\theta d\theta \Omega_{d-2}$, where the solid angle is $\Omega_{d-2} =2\pi^{(d-1)/2}/\Gamma\left((d-1)/2\right)$.

The LKF transformation that relates the coordinate space fermion propagator in Landau gauge to the one in an arbitrary covariant gauge is as follows~\cite{Landau:1955zz,Fradkin:1955jr}\,:
\begin{equation}
S_{F}\left(x;\xi\right)=S_{F}\left(x,0\right)\left[\text{e}^{i\left(\Delta_{d}\left(x\right)-\Delta_{d}\left(0\right)\right)}\right]\,,\label{eq:4}
\end{equation}
where
\begin{equation}
\Delta_{d}\left(x\right) = -ie^2\xi\mu^{4-d}\int_{0}^{\infty}\frac{d^{d}p}{\left(2\pi\right)^{d}}\frac{\text{e}^{-ip\cdot x}}{p^{4}}\,.\label{eq:5}
\end{equation}
In Eq.~(\ref{eq:5}), $e$ is the electromagnetic coupling and $\mu$ is a mass scale parameter to render $e$ dimensionless for all $d$. Performing the $d$ dimensional integration, we get \cite{Bashir:2002sp}
\begin{equation}
   \Delta_{d}\left(x\right) =-\frac{i\xi e^2}{16\left(\pi\right)^{\frac{d}{2}}}\left(\mu x\right)^{4-d}\Gamma\left(\frac{d}{2}-2\right)\,.\label{eq:6}
\end{equation}
With this generic formalism, the next step is to outline the systematic procedure to compute the results for the fermion propagator in the momentum space at any required order on applying the LKF transformations.

\subsection{Outlining the Procedure}

In principle, the intended approach is straightforward and has been explained in several works before: 
(i) We take the starting gauge as the Landau gauge and use as input both the momentum space functions $F\left(p;0\right)$ and $\mathcal{M}\left(p;0\right)$ at the lowest (tree) level in perturbation theory. One might argue that the tree level expression for the fermion propagator does not depend on the parameter $\xi$. Therefore, we could have identified these functions with any given gauge. However, as we explained before, for massless QED in arbitrary space-time dimensions $d$,  even one-loop contribution to $F\left(p;\xi\right)$ strictly vanishes in the Landau gauge~\cite{Davydychev:2000rt}.
Therefore, the Landau gauge has a definite quantitative advantage as the starting gauge corresponding to the tree level expression for the fermion propagator. (ii) We then Fourier transform the tree level fermion propagator to the coordinate space to find $X\left(x;0\right)$ and $Y\left(x;0\right)$. (iii) Subsequently, we utilize the LKF transformations to find $X\left(x;\xi\right)$ and $Y\left(x;\xi\right)$. (iv) Finally, we Fourier transform the results back to the momentum space to find $F\left(p;\xi\right)$ and $\mathcal{M}\left(p;\xi\right)$ at a given order in the coupling $e$ or $\alpha= e^2/(4 \pi)$. Employing this procedure, we obtain the two-loop results for the fermion propagator for massless and massive casess in QED3 as well as QED4.
However, before we embark upon the task, we provide a quick reference to what has been reported in the literature to date. \\

\noindent
 {\bf QED3:} Building upon some earlier related efforts in the field,~\cite{Bashir:2000rv, Bashir:1999bd, Bashir:1999xs}, the LKF transformed result for the wavefunction renormalization  $F\left(p;\xi\right)$ in massless QED3 till the two-loop order was first reported in~\cite{Bashir:2000ur}. However, for massive fermions, only the one-loop results for  $F\left(p;\xi\right)$ and the mass function $\mathcal{M}(p)$ were obtained through LKF transformation for QED3 in~\cite{Bashir:2002sp} which were then explicitly compared with the known one-loop results. Perfect agreement was found except for a gauge parameter-independent term, which is naturally allowed by the structure of the LKF transformation. \\

 \noindent
{\bf QED4:} 
For massless QED4, it has been known for a while that the LKF transformation yields a power law for $F\left(p;\xi\right)$,~\cite{Curtis:1990zs}. It has also been shown that $F\left(p;\xi\right)$ retains its power law structure even if we start from the true all order expansion of $F\left(p;0\right)$.
  However, the exponent now includes contributions from next-to-leading logarithms and so on~\cite{Bashir:1997qt}. In fact, this observation has widely been used to construct, constrain and improve  fermion-photon vertices in the field. The massive case was presented 
  in~\cite{Bashir:2002sp} and its one-loop expansion was compared with earlier results.
  
In this work for QED4, we find that at  order $\alpha^2$, we obtain both leading and next-to-leading logarithmic terms as expected. This series can again be summed up in the form of a power law structure.  For the massive case, the mass function $\mathcal{M}\left(p;\xi\right)$ and the fermion wavefunction renormalization $F\left(p;\xi\right)$ are obtained in terms of the Hypergeometric functions. For small momenta, expanding these functions in the leading powers of $p^2/m^2$ and next-to-leading order in $\alpha$, we obtain their analytical expressions. Once again, we observe the leading and next-to-leading logarithms in $F\left(p;\xi\right)$ at the two-loop levels. Till the one-loop level, we obtain the one-loop massive case results reported~\cite{Bashir:2002sp}, providing a confirmatory test of our computation. Unlike the massless case, the $F\left(p;\xi\right)$ no longer has a power law structure. However, in all cases, using the arguments of multiplicative renormalizability, we obtain the renormalization constants to two loops. In the next section, we provide explicit details of our calculations.

\section{Massless and Massive fermion Propagator at Two loops} \label{MLMFP}

In this section, we derive the results for massless and massive fermion propagator at two loops in QED3 and QED4. However, it is useful to begin with some notations and definitions in general $d$ dimensions.
Note that in the massless case, $\mathcal{M}\left(p;\xi\right)=0$. Therefore, from Eqs.~(\ref{eq:1}) and~(\ref{eq:2}) the respective fermion propagator in momentum and coordinate spaces takes the form
\begin{eqnarray}
S_{F}\left(p;\xi\right)&=&\frac{F\left(p;\xi\right)}{i\slashed{p}}\label{eq:7a}\;,\\
   S_{F}\left(x;\xi\right)&=&\slashed{x}X\left(x;\xi\right)\;.\label{eq:7b}
\end{eqnarray}
In the Landau gauge, {\em i.e.}, $\xi = 0$, Eq. (\ref{eq:3a}) becomes
\begin{equation}
S_{F}\left(x;0\right)=\int\frac{d^{d}p}{\left(2\pi\right)^{d}}\text{e}^{-ip\cdot x}S_{F}\left(p;0\right)\;. \label{8}
\end{equation}
Motivated by the lowest order in perturbation theory, we take $F\left(p;0\right)=1$ in Eq.~(\ref{eq:7a}), arriving at the corresponding coordinate space result from Eq~(\ref{eq:7b})\,:
\begin{equation}
X\left(x;0\right) =-\frac{i}{x^{2}}\int\frac{d^{d}p}{\left(2\pi\right)^{d}}\text{e}^{-ip\cdot x}\left(\frac{p\cdot x}{p^{2}}\right).\label{eq:9}
\end{equation}
For the massive fermion propagator, Eqs.~(\ref{eq:1}) 
and~(\ref{eq:2}) are combined through the Fourier transform as\,:
\begin{equation}
\hspace{-2mm} \slashed{x}X\left(x;\xi\right)+Y\left(x;\xi\right)=\int\frac{d^{d}p}{\left(2\pi\right)^{d}}\text{e}^{-ip\cdot x}\frac{F\left(p;\xi\right)}{i\slashed{p}-\mathcal{M}\left(p;\xi\right)}\;.\label{eq:10}
\end{equation}
At leading order in perturbation theory, in the Landau gauge, $F\left(p;0\right)=1$ and $\mathcal{M}\left(p;0\right)=m$, which implies
\begin{eqnarray}
X\left(x;0\right) &=&-\frac{i}{x^{2}}\int\frac{d^{d}p}{\left(2\pi\right)^{d}}\text{e}^{-ip\cdot x}\left(\frac{p\cdot x}{p^{2}+m^{2}}\right)\,,\label{eq:11a}\\
\nonumber \\
Y\left(x;0\right)&=&-m\int\frac{d^{d}p}{\left(2\pi\right)^{d}}\text{e}^{-ip\cdot x}\frac{1}{p^{2}+m^{2}}\,.\label{eq:11b}
\end{eqnarray}
With these starting expressions, the next step is to solve the Fourier integrals in three and four space-time dimensions; and then apply the LKF transformations to find the corresponding expressions in any covariant gauge $\xi$. For the sake of convenient flow of ideas and for not saturating the main text with an excessive number of formulae, the solutions of some of the useful integrals are given in Appendix~\ref{AppendixA}.

\subsection{Three Dimensional Case}

We further divide this discussion both for massless and massive fermions. 
Several results are already  known in the literatutre. Therefore, we shall make explicit comparison whenever possible and point out the differences if the case might be. 

\subsubsection{Massless Fermion Propagator in QED3} 
For the massles fermions, writing $p\cdot x = px\cos\theta$ in Eq.~(\ref{eq:9}), and using  Eqs.~(\ref{eq:A1},~\ref{eq:A2}), the integration over $p,\; \theta$ and $\phi$ yields\,:
\begin{equation}
 X\left(x;0\right) = -\frac{1}{4\pi x^3}\;.\label{eq:12}   
\end{equation}
To apply the LKF transformation, we need to rewrite Eq.~(\ref{eq:6}) explicitly in three dimensions, {\em i.e.}, 
\begin{equation}
 \Delta_{3}\left(x\right) =\frac{i\alpha\xi}{2}x\;.\label{eq:13}
\end{equation}
Therefore, Eq.~(\ref{eq:5}) permits us to write the fermion propagator in an arbitrary gauge as follows\,:
\begin{equation}
S_{F}\left(x;\xi\right)=S_{F}\left(x,0\right) \text{e}^{-\left(\alpha \xi/2\right)x}\;.\label{eq:14} 
\end{equation}
Executing the remaining steps of the outlined procedure, the two-loop result for the fermion propagator in this case was derived in Ref.~\cite{Bashir:2000ur}. Expanding around small $\alpha$, and keeping the terms to order $\alpha^2$, we rewrite the final result, also quoted in Ref.~\cite{Bashir:2000ur} in the momentum space for the sake of completeness:
\begin{equation}
    F\left(p;\xi\right) = 1-\frac{\alpha\xi\pi}{4p}- \frac{\left(28-3\pi^2\right)}{16}\frac{\alpha^2}{p^2}+\frac{\alpha^2\xi^2}{4p^2}\,.\label{eq:15}
\end{equation}
This result agrees in its totality  with that in Ref.~\cite{Bashir:2000ur} which in turn was found to be in agreement with the two-loop perturbative expression in the weak coupling limit up to the terms independent of the covariant gauge parameter $\xi$. \\

\subsubsection{Massive Fermion Propagator in QED3}

The massive fermion propagator in position space is defined in Eqs. (\ref{eq:11a}) and (\ref{eq:11b}). On performing the three dimensional integration, then using the result (\ref{eq:A3}) and the outline we suggested earlier and employed in several works including Ref.~\cite{Bashir:2002sp}, we can write:
\begin{eqnarray}
    X\left(x;0\right)&=&-\frac{1+mx}{4\pi x^{3}}\text{e}^{-mx}\; , \label{eq:16a} \\ \nonumber \\
    Y\left(x;0\right)&=&-\frac{m}{4\pi x}\text{e}^{-mx}\;. \label{eq:16b}
\end{eqnarray}
Employing Eqs. (\ref{eq:4}, \ref{eq:16a}, \ref{eq:16b}), the corresponding results in an arbitrary gauge $\xi$ are
\begin{eqnarray}
    X\left(x;\xi\right)&=&-\frac{1}{4\pi x^{3}}\left(1+mx\right)\text{e}^{-\left(m+\alpha\xi/2\right)x}\;,\label{eq:17a} \\ \nonumber \\ 
    Y\left(x;\xi\right) & = & -\frac{m}{4\pi x}\text{e}^{-\left(m+\alpha\xi/2\right)x}\;.\label{eq:17b}
\end{eqnarray}
To transform these equations back to momentum space, and using Eq. (\ref{eq:1}), we get
\begin{eqnarray}
 A\left(p;\xi\right)&=& \int d^3x\;\text{e}^{ip\cdot x}\;Y\left(x;\xi\right)\;,\label{eq:18a}\\  
  B\left(p;\xi\right)&=& \int d^3x\; p\cdot x\; \text{e}^{ip\cdot x}X\left(x;\xi\right)\;. \label{eq:18b}
\end{eqnarray}
In terms of $A\left(p;\xi\right)$ and $B\left(p;\xi\right)$ the mass function and the wavefunction can be written as:
\begin{eqnarray}
\mathcal{M}\left(p;\xi\right) &\equiv&   p^2 \, \frac{A\left(p;\xi\right)}{B\left(p;\xi\right)} \,,  \nonumber \\ F\left(p;\xi\right)&\equiv& -B\left(p;\xi\right)  -  \frac{p^{2} A^{2}\left(p;\xi\right)}{B\left(p;\xi\right)} \,.  \label{eq:18c}
\end{eqnarray}
Carrying out the angular and radial integration, the results of $ A\left(p;\xi\right)$ and $ B\left(p;\xi\right)$ involve the trigonometric functions of $p$ and $m$ too. On using these results the non-perturbative expressions for the mass function and the wave function renormalization were obtained as a function of $\xi$ in Ref.~\cite{Bashir:2002sp}. On explicit computation, we confirm that our general expression for the mass function 
\begin{equation}
 \mathcal{M}\left(p;\xi\right) =  \frac{8p^{3}m}{\Phi}\,, \label{eq:19}
\end{equation}
where 
\begin{widetext}
\begin{equation}
\Phi = 2p\left(4p^{2}+\alpha\xi\left(2m+\alpha\xi\right)\right)-\alpha\xi\left(4p^{2}+\left(2m+\alpha\xi\right)^{2}\right)\tan^{-1}\left(\frac{2p}{2m+\alpha\xi}\right)\,, \label{eq:20}   
\end{equation}
is in agreement with Eq.~(19) of 
Ref.~\cite{Bashir:2002sp}. The same is true for the result for the wave function renormalization, see Ref.~\cite{Bashir:2002sp}\,:
\begin{equation}
 F\left(p;\xi\right)= \frac{\left(4p^{2}+\alpha\xi\left(2m+\alpha\xi\right)\right)\Phi + 32m^{2}p^{3}} {\left(4p^{2}+\left(2m+\alpha\xi\right)^{2}\right)\Phi}-\frac{\alpha\xi}{2p}\tan^{-1}\left(\frac{2p}{2m+\alpha\xi\mu}\right)\,.\label{eq:21}  
\end{equation}
Extending and following the massles case of Ref.~\cite{DallOlio:2021njq},
we report the fermion propagator at the two loops in this article, \textit{i.e.}, to order $\alpha^2$; Therefore, expanding Eqs.~(\ref{eq:19}) and~(\ref{eq:21}), we arrive at these expressions\,:
\begin{eqnarray}
    \mathcal{M}\left(p;\xi\right)&=&m\left[1-\frac{\alpha\xi}{2}\left(\frac{mp-\left(p^{2}+m^{2}\right)\tan^{-1}\left(\frac{p}{m}\right)}{p^{3}}\right)\right.\notag\\
    &&\left.+\frac{\left(\alpha\xi\right)^{2}}{4}\left(\frac{p^{2}\left(m^{2}-2p^{2}\right)-2m^{3}p\tan^{-1}\left(\frac{p}{m}\right)+\left(p^{2}+m^{2}\right)^{2}\tan^{-1}\left(\frac{p}{m}\right)^{2}}{p^{6}}\right)\right]\,,\notag\\ \nonumber \\
    F\left(p;\xi\right)&=&1-\frac{\alpha\xi}{2}\left(\frac{mp+\left(p^{2}-m^{2}\right)\tan^{-1}\left(\frac{p}{m}\right)}{p^{3}}\right)\notag\\
    &&+\frac{\left(\alpha\xi\right)^{2}}{4}\left[\frac{p^{2}\left(p^{4}+m^{4}\right)-2m^{3}p\left(p^{2}+m^{2}\right)\tan^{-1}\left(\frac{p}{m}\right)+m^{2}\left(p^{2}+m^{2}\right)^{2}\tan^{-1}\left(\frac{p}{m}\right)^{2}}{p^{6}\left(p^{2}+m^{2}\right)}\right]\,.\label{eq:22}
\end{eqnarray}
\end{widetext}
In addition to comparing the results of 
Eq.~(\ref{eq:21}) with the ones reported in Ref.~\cite{Bashir:2002sp}, it is also a promising check if the two-loop results of the mass function and wavefunction renormalization written above can be retrieved by following an alternative route: we can start from the two-loop expanded form of $A\left(p;\xi\right)$ and $B\left(p;\xi\right)$ defined in Eqs.~(\ref{eq:18a}) and~(\ref{eq:18b}), respectively. We can then follow the same procedure to arrive at the two-loop results directly instead of going through the non-perturbative results first. Therefore, explicitly, we have the following two loops result \\
\vspace{-0.5cm}
\begin{equation}
 A\left(p;\xi\right)=\frac{m}{4\pi}\int d^{3}x\;\text{e}^{ip\cdot x}\frac{\text{e}^{-mx}}{x}\left[1-\frac{\alpha\xi x}{2}+\frac{\left(\alpha\xi x\right)^{2}}{8}\right] .   \nonumber
\end{equation}
Performing angular integration gives
\begin{equation}
 A\left(p;\xi\right) = -\frac{m}{p}\int dx\;\text{e}^{-mx}\sin px\left[1-\frac{\alpha\xi x}{2}+\frac{\left(\alpha\xi x\right)^{2}}{8}\right]\,,\notag   
\end{equation}
and the radial integration, on using the standard integrals, yields
\begin{widetext}
\begin{equation}
    A\left(p;\xi\right)=-\frac{m}{p^{2}+m^{2}}\left[1-\frac{\alpha\xi}{2}\frac{2m}{p^{2}+m^{2}}-\frac{\left(\alpha\xi\right)^{2}}{4}\frac{p^{2}-3m^{2}}{\left(p^{2}+m^{2}\right)^{2}}\right]\,.\label{eq:23}
\end{equation}
Following the same steps used above, with the help of integrals given in 
Eq.\,(\ref{eq:A4}), we obtain
    \begin{equation}
    B\left(p;\xi\right)=-\frac{p^{2}}{p^{2}+m^{2}}\left[1-\frac{\alpha\xi}{2}\left(\frac{m\left(p^{2}-m^{2}\right)}{p^{2}\left(p^{2}+m^{2}\right)}+\frac{p^{2}+m^{2}}{p^{3}}\tan^{-1}\left(\frac{p}{m}\right)\right)
    +\frac{\left(\alpha\xi\right)^{2}}{4}\frac{\left(p^{2}+5m^{2}\right)}{\left(p^{2}+m^{2}\right)^{2}}\right]\,.\label{eq:24}
\end{equation}
\end{widetext}
Finally, by substituting these functions in Eq.~(\ref{eq:18c})
and retaining the terms to two loops, {\em i.e.}, till ${\cal O}(\alpha^2)$, we can readily confirm that that the result presented in Eq.~(\ref{eq:22}) is faithfully reproduced. This test authenticate that our generic expression given in Eq.~(\ref{eq:21}) is indeed correct.
We now discuss and analyze the 4-dimensional case.

\subsection{Four Dimensional Case}
The method for computing the fermion propagator in QED4 is the same as that in the QED3, \textit{i.e.,} starting with Landau gauge, we have to find $X\left(x;\xi\right)$ and $Y\left(x;\xi\right)$, and then fourier transform them back to momentum space. Again, we start with the massless case in QED4 and find the fermion propagator to the two-loops order. Later, we include the mass term to calculate the two-loop results for the massive fermion.

\subsubsection{Massless Fermion Propagator in QED4}

Employing Eqs.\,(\ref{eq:11a}) and\,(\ref{eq:11b}), the lowest order fermion propagator in position space becomes
\begin{equation}
X\left(x;0\right)=-\frac{1}{2\pi^{2}x^{4}}\,, \quad Y\left(x;0\right)=0\,.\label{eq:25}
\end{equation}
Multiplying it with $\slashed{x}$, the corresponding fermion propagator in the coordinate space is straightforwardly\,:
\begin{equation}
S_{F}\left(x;0\right)=\slashed{x}X\left(x;0\right)=-\slashed{x}\left(\frac{1}{2\pi^{2}x^{4}}\right) \,.\label{eq:26}
\end{equation}
To apply the LKF transformation in 4-dimensions while avoiding the singularity arising due to $\Gamma(\varepsilon \rightarrow 0)$, we need to expand Eq.\,(\ref{eq:6}) around $d=4-\varepsilon$, which yields
\begin{equation}
\Delta_{4-\varepsilon}\left(x\right)  =\frac{-i \alpha \xi}{4\left(\pi\right)^{1-\frac{\varepsilon}{2}}}\left(\mu x\right)^{\varepsilon}\left[-\frac{2}{\varepsilon}-\gamma_E+\mathcal{O}\left(\varepsilon\right)\right],\label{eq:27}
\end{equation}
where $\mu$ is a mass scale parameter to keep the coupling constant dimensionless while $\gamma_E$ is Euler-Mascheroni constant. Using the identity $ \left(\mu x\right)^{\varepsilon} = 1+\varepsilon\left(\mu x\right)+\mathcal{O}\left(\varepsilon^2\right)$, and working in the 
$\overline{\text{MS}}$ scheme, the above result reduces to
\begin{equation}
i\Delta_{4-\varepsilon}\left(x\right)=-\frac{\alpha \xi }{4\pi}\log\left(\mu^{2}x^{2}\right)\,.\label{eq:28}
\end{equation}
Now when $x \rightarrow 0$, $\Delta_{F}\left( x \rightarrow  0\right)$ becomes divergent. Therefore, in order to cure this divergence, one possibility is to define a position space cut-off $\left(x_{\text{min}}\right)$. It allows us to rewrite $\Delta_{4-\varepsilon}$ as
\begin{equation}
i\Delta_{4-\varepsilon}\left(x_{\text{min}}\right)=-\frac{\alpha \xi }{4 \pi}\log\left(\mu^{2}x_{\text{min}}^{2}\right)\,.\label{eq:29}
\end{equation}
Thus, to the lowest order, we have
\begin{equation}
i\Delta_{4-\varepsilon}\left(x\right)-i\Delta_{4-\varepsilon}\left(x_{\text{min}}\right)=-\frac{\alpha \xi}{4 \pi}\log\left(\frac{x^{2}}{x_{\text{min}}^{2}}\right)\,.\label{eq:30}
\end{equation}
Making use of this result along with the LKF transformation, Eq.~(\ref{eq:25}), we can write
\begin{equation}
X\left(x;\xi\right) =-\frac{1}{2\pi^{2}x^{4}}\left(\frac{x^{2}}{x_{\text{min}}^{2}}\right)^{-\nu}\,,\label{eq:31}
\end{equation}
where $\nu=\alpha\xi/4\pi$.
To transform them back to the momentum space, we use the relations given by Eqs.~(\ref{eq:18a}) and~(\ref{eq:18b}), arriving explicitly at\,:
\begin{eqnarray}
A\left(p;\xi\right)&=& 0 \,,\notag\\
iB\left(p;\xi\right)&=&\int d^{4}x\left(p\cdot x\right)\text{e}^{ip\cdot x}X\left(x;\xi\right)\,.\label{eq:32}
\end{eqnarray}
Carrying out the angular integration in 4-dimensional Euclidean space and employing Eq.~(\ref{eq:31}), we obtain
\begin{equation}
B\left(p;\xi\right) =-\frac{2}{\left(x_{\text{min}}\right)^{-2\nu}}\int dx \left(x\right)^{-2\nu-1}J_{2}\left(px\right)\,,\label{eq:33}
\end{equation}
where $J_2\left(px\right)$ is the Bessel function of first-kind. 
Now in order to perform the radial integration, we make use of the following identity:
\begin{equation}
\int dxx^{\alpha}J_{\beta}\left(bx\right)  =2^{\alpha}\frac{\Gamma\left(\frac{1+\alpha+\beta}{2}\right)}{\Gamma\left(\frac{1-\alpha+\beta}{2}\right)}\,,\label{eq:34}
\end{equation}
in Eq.~(\ref{eq:33}) to rewrite it as
\begin{equation}
B\left(p;\xi\right) =-\frac{2}{\left(px_{\text{min}}\right)^{-2\nu}}\frac{1}{2^{2\nu +1}}\frac{\Gamma\left(1-\nu\right)}{\Gamma\left(2+\nu\right)}\,.\label{eq:35}
\end{equation}
From Eq.~(\ref{eq:18c}), the corresponding wavefunction renormarlization in the massless case becomes~\cite{Aslam:2015nia}\,:
\begin{equation}
F\left(p;\xi\right) =\frac{2}{\left(px_{\text{min}}\right)^{-2\nu}}\frac{1}{2^{2\nu +1}}\frac{\Gamma\left(1-\nu\right)}{\Gamma\left(2+\nu\right)}\,.\label{eq:36}
\end{equation}
This is a general result which is valid to all orders in $\alpha$. 
As our purpose here is to study the fermion propagator at two loops, 
we expand it out to ${{\cal O}(\nu^2)} = {{\cal O}(\alpha^2)} $\,: 
\begin{eqnarray}
&& \hspace{-6mm} \left(\frac{px_{\text{min}}}{2}\right)^{2\nu}  = 1+2\nu \log\left(\frac{px_{\text{min}}}{2}\right)+2\nu^{2}\log^{2}\left(\frac{px_{\text{min}}}{2}\right) \,,\notag \\
&& \hspace{-6mm} \frac{\Gamma\left(1-\nu \right)}{\Gamma\left(2+\nu \right)}  =1+\left(2\gamma_E-1\right)\nu
+\left(1-2\gamma_E+2\gamma_E^{2}\right)\nu^{2}\,.\label{eq:37}
\end{eqnarray}
Finally, on using Eq.~(\ref{eq:37}) in 
Eq.~(\ref{eq:36}), and again keeping the terms to the order ${{\cal O} {(\nu^2)} }\equiv {{\cal O} {(\alpha^2)}}$, we have
\begin{eqnarray}
F\left(p;\xi\right)&=&1+\frac{\xi\alpha}{4\pi}\left(2\gamma_E-1+\log\left(\frac{p^2}{\Lambda^2}\right)\right)\notag\\
&+&\left(\frac{\xi\alpha}{4\pi}\right)^{2}\bigg(1-2\gamma_E+2\gamma_E^{2}\notag\\
&+&\left(2\gamma_E-1\right)\log\left(\frac{p^2}{\Lambda^2}\right)+\frac{1}{2}\log^{2}\left(\frac{p^2}{\Lambda^2}\right)\bigg) \,, \qquad \label{eq:38}
\end{eqnarray}
with $x_{\text{min}}= 2/\Lambda$. 
We shall discuss the renormalization of our two-loops regularized results and its comparison with the QED inferred result of Ref.~\cite{DallOlio:2021njq}
in Sect.~\ref{renormalization}.

\subsubsection{Massive Fermion Propagator in QED4}

We start by recalling that the LKF transformed massive fermion propagator at one-loop is derived in~\cite{Bashir:2002sp}. In order to extend it to the two-loop level, we again start with lowest order mass function and the wavefunction renormalization in the Landau gauge, that is to say, 
 $F\left(p;0\right) = 1, \,\mathcal{M}\left(p;0\right) = m$.
Substituting these expressions in Eqs.~(\ref{eq:11a}) and (\ref{eq:11b}), we obtain the corresponding fermion propagator at the lowest order in this particular gauge in the coordinate space. The angular integration yields the known results\,:
\begin{eqnarray}
X\left(x;0\right)&=&-\frac{1}{4\pi^{2}x^{2}}\int_{0}^{\infty}dp\frac{p^{3}}{p^{2}+m^{2}}J_{2}\left(px\right)\,,\label{eq:39a} \\ \nonumber \\
Y\left(x;0\right) &=& -\frac{m}{4\pi^{2}x}\int_{0}^{\infty}dp\frac{p^{2}}{p^{2}+m^{2}}J_{1}\left(px\right)\,.\label{eq:39b}
\end{eqnarray}
In order to perform the radial integration, we use 
\begin{equation}
\int_{0}^{\infty}dp\frac{p^{\beta+1}}{\left(p^{2}+m^{2}\right)^{\mu+1}}J_{\beta}\left(px\right)=\frac{m^{\beta-\mu}x^{\mu}}{2^{\mu}\Gamma\left(\mu+1\right)}K_{\beta-\mu}\left(mx\right)\;.\label{eq:40}
\end{equation}
In our case, $\mu=0$ and $\beta=2$ and $1$ in Eqs.~(\ref{eq:39a}) 
and~(\ref{eq:39b}), respectively. Therefore,
\begin{eqnarray}
X\left(x;0\right)&=&-\frac{m^{2}}{4\pi^{2}x^{2}}K_{2}\left(mx\right)\,,\label{eq:41a}\\ \nonumber \\
Y\left(x;0\right)&=&-\frac{m^{2}}{4\pi^{2}x}K_{1}\left(mx\right)\,,\label{eq:41b}
\end{eqnarray}
where $K_1$ and $K_2$ are the modified Bessel functions of second kind. 
Together with Eq.~(\ref{eq:30}), the LKF transformed versions of Eqs.~(\ref{eq:41a}) and (\ref{eq:41b}) are\,:
\begin{eqnarray}
X\left(x;\xi\right) &=&-\frac{m^{2}}{4\pi^{2}x^{2}}K_{2}\left(mx\right)\left(\frac{x^{2}}{x_{\text{min}}^{2}}\right)^{-\nu}\,,\label{eq:42}\\ \nonumber \\
Y\left(x;\xi\right) & = &-\frac{m^{2}}{4\pi^{2}x}K_{1}\left(mx\right)\left(\frac{x^{2}}{x_{\text{min}}^{2}}\right)^{-\nu}\,.\label{eq:43}
\end{eqnarray}
Again, transforming them back to momentum space through Eqs.~(\ref{eq:18a}) and~(\ref{eq:18b}), and performing the 4-dimensional angular integration, we obtain\,:
\begin{eqnarray}
&& \hspace{-10mm} A\left(p;\xi\right)= -\frac{m^{2}}{p} \int dx\;xJ_{1}\left(px\right)K_{1}\left(mx\right)\left(\frac{x^{2}}{x_{\text{min}}^{2}}\right)^{-\nu},\notag\\ \nonumber \\
&&  \hspace{-10mm} B\left(p;\xi\right)= -m^{2}\int dx\;xJ_{2}\left(px\right)K_{2}\left(mx\right)\left(\frac{x^{2}}{x_{\text{min}}^{2}}\right)^{-\nu}.
\label{eq:45}
\end{eqnarray}
To solve these radial integrals, one can use the result of the standard integral
\begin{eqnarray}
\int dxx^{-\lambda}J_{\beta}\left(bx\right)K_{\mu}\left(ax\right) &=&
 \frac{a^{\lambda-\beta-1}b^{\beta}}{2^{\lambda+1}\Gamma\left(1+\beta\right)}\Gamma\left( a \right)\Gamma\left( b \right)\notag\\ 
&\times&
{}_{2}F_{1}\left(a,b;c;z\right)\;,\label{eq:46}
\end{eqnarray}
where $_{2}F_{1}\left(a,b;c;z\right)={}_{2}F_{1}\left(b,a;c;z\right)$ are the well-known confluent Hyper-geometric functions. In our case, the arguments of these Hyper-geometric functions are defined through the following relations\,:
\begin{eqnarray}
a &=& \frac{\beta-\lambda+\mu+1}{2}\,, \nonumber  \\
b  &=&  \frac{\beta-\lambda-\mu+1}{2}\,,\notag\\
c &=&  \beta+1\,,  \nonumber \\ 
z &=& -\frac{b^{2}}{a^{2}}\,. \label{eq:47} \\ \nonumber
\end{eqnarray}
Now, in the case of $A\left(p;\xi\right)$ in Eq.~(\ref{eq:45}), we have the values $\beta=1, \, \lambda=2\nu-1, \, \mu=1,\, a=m$ and $b=p$. Therefore, 
\begin{eqnarray}
A\left(p;\xi\right) &=& -\frac{1}{m}\left(\frac{m^{2}}{\Lambda^{2}}\right)^{\nu}\Gamma\left(2-\nu\right)\Gamma\left(1-\nu\right)\times\notag\\
&&{}_{2}F_{1}\left(2-\nu,1-\nu;2;-\frac{p^{2}}{m^{2}}\right)\;. \label{eq:48}
\end{eqnarray}
Similarly, for $B\left(p;\xi\right)$, we make use of the fact that we have $\beta=\mu=2,\, a=m,\, b=p$ and $\lambda=2\nu-1\,$ in Eq.~(\ref{eq:48}). Therefore, we obtain\,:
\begin{eqnarray}
B\left(p;\xi\right)&=&-\frac{p^{2}}{2m^{2}}\left(\frac{m^{2}}{\Lambda^{2}}\right)^{\nu}\Gamma\left(3-\nu\right)\Gamma\left(1-\nu\right)\times\notag\\
&&{}_{2}F_{1}\left(3-\nu,1-\nu;3;-\frac{p^{2}}{m^{2}}\right).\label{eq:49}
\end{eqnarray}
The use of these explicit expressions for $A\left(p;\xi\right)$ and $B\left(p;\xi\right)$ in Eq.~(\ref{eq:18c}) implies
\begin{widetext}
\begin{eqnarray}
F\left(p;\xi\right)&=&\left(\frac{m^{2}}{\Lambda^{2}}\right)^{\nu}\frac{1}{2m^{2}\Gamma\left(3-\nu\right)}\frac{\Gamma\left(2-\nu\right)\Gamma\left(1-\nu\right)}{_{2}F_{1}\left(3-\nu,1-\nu;3;-\frac{p^{2}}{m^{2}}\right)}
\left\{ p^{2}\frac{\Gamma^{2}\left(3-\nu\right)}{\Gamma\left(2-\nu\right)}{}_{2}F_{1}^{2}\left(3-\nu,1-\nu;3;-\frac{p^{2}}{m^{2}}\right)\right.\nonumber \\
 & & \left.+\;4m^{2}\Gamma\left(2-\nu\right){}_{2}F_{1}^{2}\left(2-\nu,1-\nu;2;-\frac{p^{2}}{m^{2}}\right)\right\}\;, \label{eq:50}
\end{eqnarray}
and
\begin{equation}
\mathcal{M}\left(p;\xi\right)  = \frac{2m}{\left(2-\nu\right)}\frac{_{2}F_{1}\left(2-\nu,1-\nu;2;-\frac{p^{2}}{m^{2}}\right)}{_{2}F_{1}\left(3-\nu,1-\nu;3;-\frac{p^{2}}{m^{2}}\right)}\;.\label{eq:51}
\end{equation}
\end{widetext}
 These results, although written more compactly, are in agreement with~\cite{Bashir:2002sp}. It is important to mention here that these results derived through the LKF transformation contain information of higher orders too, in further sections we will explore their particularity at two-loop level for some special cases.
For $\alpha=0$, employing the identities of the Hyper-geometric equations, it is easy to infer that
 the leading order result for the electron propagator is recovered\,:  
\begin{equation}
S_{F}\left(p;\xi\right)=\frac{F\left(p;\xi\right)}{\slashed{p}-\mathcal{M}\left(p;\xi\right)}=\frac{1}{\slashed{p}-m}\; .\label{eq:56}
\end{equation}
We look at the two-loop case in the limit $m\gg p$.
In this limit, the Hyper-geometric function can be expanded
in the powers of $\frac{p^{2}}{m^{2}}$ as~\cite{Bashir:2002sp}\,:
\begin{equation}
_{2}F_{1}\left(\alpha,\beta;\gamma_E;-\frac{p^{2}}{m^{2}}\right)=1-\frac{\alpha\beta}{\gamma_E}\frac{p^{2}}{m^{2}}+\mathcal{O}\left(\frac{p^{2}}{m^{2}}\right)^{2}\;,\label{eq:57}
\end{equation}
which gives us
\begin{eqnarray}
_{2}F_{1}\left(2-\nu,1-\nu;2;-\frac{p^{2}}{m^{2}}\right)&=&  1-\frac{\left(2-\nu\right)\left(1-\nu\right)}{2}\frac{p^{2}}{m^{2}}\;,\notag\\
_{2}F_{1}\left(3-\nu,1-\nu;3;-\frac{p^{2}}{m^{2}}\right)&=&  1-\frac{\left(3-\nu\right)\left(1-\nu\right)}{3}\frac{p^{2}}{m^{2}}\;.\notag
\end{eqnarray}
Employing these expansion, the two-loop result for the mass function reads as\,:
\begin{widetext}
\begin{equation}
\mathcal{M}\left(p;\xi\right)=m\left[1+\frac{1}{2}\frac{\alpha\xi}{4\pi}\left(1+\frac{1}{3}\frac{p^{2}}{m^{2}}\right)+\frac{1}{4}\left(\frac{\alpha\xi}{4\pi}\right)^{2}\left(1-\frac{1}{3}\frac{p^{2}}{m^{2}}\right)\right] \,.\label{eq:58}
\end{equation}
If we keep the leading order terms in $\nu$, it can be seen that the result of one-loop mass function agrees with Eq.~(41) in Ref.~\cite{Bashir:2002sp}.
Now expanding out till the second order terms in $\nu$ and leading order in $p^2/m^2$ in $F\left(p;\xi\right)$, we get:
\begin{eqnarray}
F\left(p;\xi\right) & =&1+\frac{\xi\alpha}{4\pi}\left(2\gamma_E-\frac{1}{2}+\log\left(\frac{m^{2}}{\Lambda^{2}}\right)+\frac{2}{3}\frac{p^{2}}{m^{2}}\right)
 +\left(\frac{\xi\alpha}{4\pi}\right)^{2}\left[2\gamma_E^{2}-\gamma_E-\frac{1}{4}+\frac{\pi^{2}}{6}+\left(2\gamma_E-\frac{1}{2}\right)\log\left(\frac{m^{2}}{\Lambda^{2}}\right)\right.\notag\\
 &&\left.+\frac{1}{2}\log^{2}\left(\frac{m^{2}}{\Lambda^{2}}\right)+\left(\frac{p^{2}}{m^{2}}\right)\left(\frac{4}{3}\gamma_E-\frac{3}{4}+\frac{2}{3}\log\left(\frac{m^{2}}{\Lambda^{2}}\right)\right)\right]\; .\label{eq:59}
\end{eqnarray}
\end{widetext}
Again, it reproduces the one-loop wave function results given in 
Ref.~\cite{Bashir:2002sp} (c.f. 
Eq.~(40)) which is obtained by dropping the terms proportional to $\nu^2$. Noticeably the one-loop is straightforwardly multiplicatively renormalizable for $\mathcal{M}\left(p;\xi\right)$ and $F\left(p;\xi\right)$. The renormalization of the two-loop results will be discussed in the next section. 

\section{Renormalization of two-loop results in QED4}\label{renormalization}

QED is a renormalized theory to every order in the perturbation theory. 
In this section, we shall provide the renormalized results for the wavefunction renormalization and the mass function at the two-loops level both for the massless and massive fermions. Let us recall
\begin{equation}
\frac{F_{R}\left(p^2/\mu^2;\xi\right)}{F\left(p^2/\Lambda^2;\xi\right)} = \mathcal{Z}^{-1}_2\left(\mu^2;\Lambda^2\right)
\,.\label{eq:60}
\end{equation}
We have used equivalent but conceptually clearer notation $F\left(p^2/\Lambda^2;\xi\right)\equiv F\left(p;\xi\right)$, which is given in Eqs.~(\ref{eq:38}) and~(\ref{eq:59}) for the massless and massive fermions, respectively. $\mathcal{Z}^{-1}_2\left(\mu^2;\Lambda^2\right)$ is the renormalization constant for fermion propagator at some renormalization scale $\mu^2$, and the purpose here is to extract it at the two loops in QED4. Till the two-loops order, {\em i.e.,} ${\cal {O}}(\nu^2)={\cal {O}}(\alpha^2)$,  we can now write\,: \vspace{-5mm}
\begin{equation}
\mathcal{Z}^{-1}_2\left(\mu^2;\Lambda^2\right) = 1+\nu\mathcal{Z}^{-1}_{2;1}+\nu^2\mathcal{Z}^{-1}_{2;2} \;,\label{eq:61}
\vspace{1mm}
\end{equation}
where we have used the notation $\mathcal{Z}^{-1}_{2;1}$ and $\mathcal{Z}^{-1}_{2;2}$ for the first and second order values of the renormalization constant
$\mathcal{Z}^{-1}_2$. The additional subscripts correspond to the number of loops. We have also taken the liberty not to show the $\mu$-dependence 
explicitly on the right hand side of the equation for the expressions not to appear very cumbersome. Moreover, notice that the usage of Eq.~(\ref{eq:61}) in the renormalization condition of Eq.~(\ref{eq:60}) implies\,:
\begin{equation}
F_{R}\left(p^2/\mu^2;\xi\right)  = \left(1+\nu \mathcal{Z}^{-1}_{2;1}+\nu^{2}\mathcal{Z}^{-1}_{2;2}\right)F\left(p^2/\Lambda^2;\xi\right)\;.\label{eq:rcondition}
\end{equation}
For the massless fermions, from Eq. (\ref{eq:38}), we can explicitly write
\begin{widetext}
\begin{eqnarray}
 F_{R}\left(p^2/\mu^2;\xi\right) & = & 1+\nu\left(2\gamma_{E}-1+\log\left(\frac{p^{2}}{\Lambda^{2}}\right)+\mathcal{Z}^{-1}_{2;1}\right)+\nu^{2}\mathcal{Z}^{-1}_{2;2}+\nu^2\left(2\gamma_{E}-1+\log\left(\frac{p^{2}}{\Lambda^{2}}\right)\right)\mathcal{Z}^{-1}_{2;1}\notag\\
   &&+\nu^{2}\left[1-2\gamma_{E}+2\gamma_{E}^{2}+\left(2\gamma_{E}-1\right)\log\left(\frac{p^{2}}{\Lambda^{2}}\right)+\frac{1}{2}\log^{2}\left(\frac{p^{2}}{\Lambda^{2}}\right)\right]\;.\label{eq:62}
\end{eqnarray}
Collecting the divergent terms from the above equation at both orders of perturbation theory, we have
\begin{eqnarray}
\mathcal{O}\left(\nu\right) & :& \mathcal{Z}^{-1}_{2;1}+\log\left(\frac{p^{2}}{\Lambda^{2}}\right)\,,\label{eq:63} \\
\mathcal{O}\left(\nu^{2}\right)  & :& \mathcal{Z}^{-1}_{2;2}+\left(2\gamma_{E}-1+\log\left(\frac{p^{2}}{\Lambda^{2}}\right)\right)\mathcal{Z}^{-1}_{2;1}+\left(2\gamma_{E}-1\right)\log\left(\frac{p^{2}}{\Lambda^{2}}\right)+\frac{1}{2}\log^{2}\left(\frac{p^{2}}{\Lambda^{2}}\right) \,.\label{eq:64}
\end{eqnarray}
\end{widetext}
Therefore, the finiteness of the renormalized result to the order ${\cal {O}}(\nu)={\cal {O}}(\alpha)$  requires\,:
\begin{equation}
\mathcal{Z}^{-1}_{2;1}=-\log\left(\frac{\mu^{2}}{\Lambda^{2}}\right)\,.\label{eq:65}
\end{equation}
Using this result in Eq.~(\ref{eq:64}), we can see that the order ${\cal {O}}(\nu^2)={\cal {O}}(\alpha^2)$ 
 term reduces to
\begin{eqnarray}
 \mathcal{O}\left(\nu^{2}\right) & :& \mathcal{Z}^{-1}_{2;2}+\left(2\gamma_{E}-1\right)\log\left(\frac{p^{2}}{\mu^{2}}\right)\notag\\
 &&+\log\left(\frac{p^{2}}{\Lambda^{2}}\right)\left[\frac{1}{2}\log\left(\frac{p^{2}}{\mu^{2}}\right)-\frac{1}{2}\log\left(\frac{\mu^{2}}{\Lambda^{2}}\right)\right]\,,\notag
\end{eqnarray}
which, for having the finite  renormalized result at the two-loops level, requires\,:
\begin{equation}
\mathcal{Z}^{-1}_{2;2} =\frac{1}{2}\log^{2}\left(\frac{\mu^{2}}{\Lambda^{2}}\right). \label{eq:66}
\end{equation}
Finally, using these extracted values of the  renormalization constant coefficients, the renormalization constant $\mathcal{Z}^{-1}_2$, defined through Eq.~(\ref{eq:61}) for the massless fermions till two loops, takes the form\,:
\begin{equation}
\mathcal{Z}^{-1}_2\left(\mu^2;\Lambda^2\right) = 1-\frac{\alpha\xi}{4\pi}\log\left(\frac{\mu^{2}}{\Lambda^{2}}\right)+\frac{1}{2}\left(\frac{\alpha\xi}{4\pi}\right)^2\log^{2}\left(\frac{\mu^{2}}{\Lambda^{2}}\right)\,. \label{eq:mlZ2}
\end{equation}
In Eq.~(30) of the Ref.~\cite{DallOlio:2021njq}, the renormalization constant for the fermion propagator at two loops was calculated through the generalized transformations in  QCD. By setting the Casimir operators $C_A =0$ and $C_F=1$, one can retrieve the corresponding QED result, and the comparison shows that the two results coincide exactly.  
We can now follow the same procedure for massive fermions in QED and define
\vspace{-1mm}
\begin{eqnarray} 
\frac{F_{R}\left(p^2/m^2,m^2/\mu^2;\xi\right)}{F\left(p^2/m^2,m^2/\Lambda^2;\xi\right)} = \mathcal{Z}^{-1}_2\left(\mu^2;\Lambda^2\right)\,,\label{eq:67} \\ \nonumber
\end{eqnarray}
where we have again tinkered with the notation in an obvious and convenient manner.
Using Eq.~(\ref{eq:59}) and the massive version of Eq.~(\ref{eq:rcondition}), we arrive at the the following result for the renormalized fermion propagator for the massive fermions till the two-loops order\,:
\begin{widetext}
\begin{eqnarray}
F_{R}\left(p,m^2/\mu^2;\xi\right) & = & 1+\frac{\xi\alpha}{4\pi}\left(2\gamma_E-\frac{1}{2}+\log\left(\frac{m^{2}}{\mu^{2}}\right)+\frac{2}{3}\frac{p^{2}}{m^{2}}\right)
 +\left(\frac{\xi\alpha}{4\pi}\right)^{2}\left[2\gamma_E^{2}-\gamma_E-\frac{1}{4}+\frac{\pi^{2}}{6}\right.\notag\\
 &&\left.+\left(2\gamma_E-\frac{1}{2}\right)\log\left(\frac{m^{2}}{\mu^{2}}\right)+\frac{1}{2}\log^{2}\left(\frac{m^{2}}{\mu^{2}}\right)+\left(\frac{p^{2}}{m^{2}}\right)\left(\frac{4}{3}\gamma_E-\frac{3}{4}+\frac{2}{3}\log\left(\frac{m^{2}}{\mu^{2}}\right)\right)\right]\, .\label{eq:68}
\end{eqnarray}
\end{widetext}
Note that we could not compare this final results with Ref.~\cite{DallOlio:2021njq} as the latter only studied massless fermion. 
This section completes our analysis of the LKF transformations in the perturbative domain both for the massless and massive fermions in QED3 and QED4. It encompasses all earlier works on the subject and extends the results to two-loops, including multiplicative renormalization of the regurlarized expressions in QED4. We present all analysis in the Euclidean space. For its correspondence with the Minkowski space on Wick rotation, we refer the reader to Refs.~\cite{James:2019ctc,Teber:2018goo}.   

As we have mentioned earlier in the text, fermion masses can be dynamically generated through chiral symmetry breaking. In the next section, we study the implications of local gauge transformations for the dynamically generated fermion propagator, in particular the fermion mass function, the chiral fermion condensate and the Euclidean pole mass for a representative functional form of the solution in the Landau gauge.  
\vspace{-0.5 cm}

\section{Gauge dependence of the dynamically generated mass} \label{DGM}

\begin{figure}[h]
\begin{centering}
\includegraphics[scale=0.9]{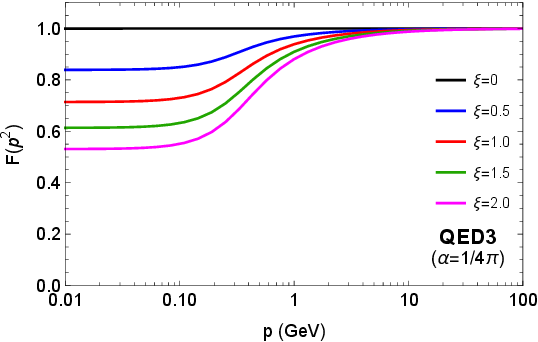}
\includegraphics[scale=0.9]{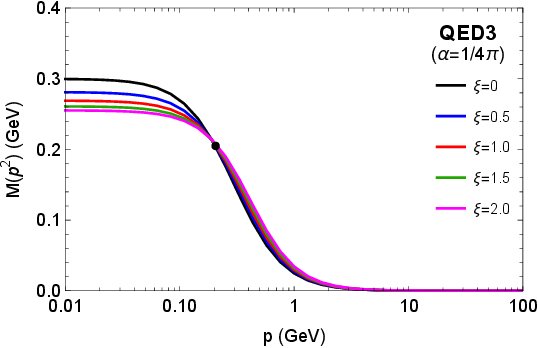}
\par\end{centering}
\caption{Wavefunction renormalization functions (above) and mass functions (below)
for various values of gauge parameter $\xi$. The black solid blob in the plots
of mass function represents the virtually constant value of the Euclidean pole
mass. We take $\alpha=1/4\pi$ (with mass dimensions in GeV) as in Ref.~\cite{Bashir:2005wt}.}\label{Fig1}
\end{figure}

\begin{figure}[h]
\begin{centering}
\includegraphics[scale=0.9]{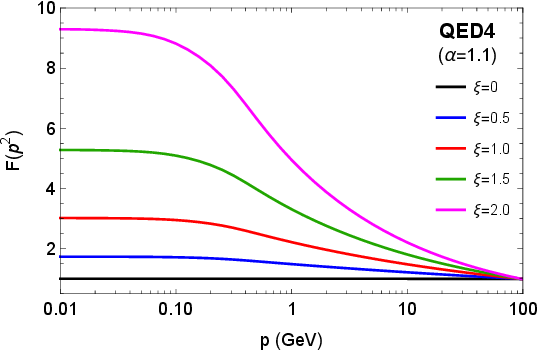}
\includegraphics[scale=0.9]{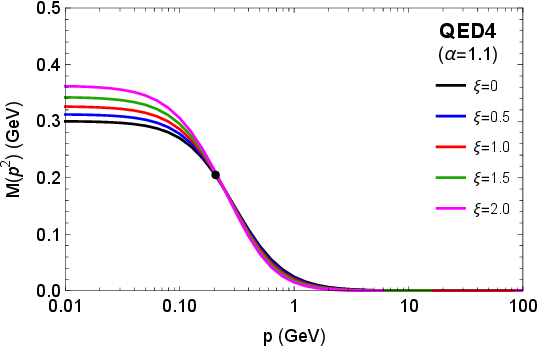}
\par\end{centering}
\caption{Wavefunction renormalization functions (above) and mass functions (below)
for various values of the gauge parameter $\xi$. The black blob in the plots
of mass functions represents the practically constant value of the Euclidean pole
mass. We take $\alpha=1.1$ and cut-off $\Lambda=10^{2}$ GeV.}\label{Fig2}
\end{figure}

In this section, we apply the LKF transformation on a fermion propagator
that has known infrared enhanced plateau for the mass function corresponding to
dynamically generated mass and falls-of as $1/p^{2}$ for large momenta. The wavefunction renormalization and the mass
functions in this representative solution in Landau gauge can be represented reasonably adequately as follows\,:
\[
F(p;0)=1,\quad\mathcal{M}(p;0)=\frac{m^{3}}{p^{2}+m^{2}},
\]
where $m$ is taken as 0.3 GeV to make its qualitative connection with the dressed light quark masses in QCD. Its  value is naturally correlated with the value of the strength of the interaction coupling $\alpha$. However, our particular choice of 0.3 GeV for the representative solution in no way affects our discussion on the gauge dependence of the propagator. It can be rescaled to any other value more realistic to QED3 without affecting our conclusions.   

The Euclidean pole mass of the corresponding
dressed fermion propagator is at 0.205 GeV, obtained by solving the equation
$\mathcal{M}(p)=p$. The fermion condensate in arbitrary $d$-dimensions is given
as
\begin{eqnarray}
\left\langle \bar{\psi}\psi\right\rangle _{\xi=0}^{d} & = & -\text{Tr}_{\mathrm{D}}[S(p;0)] \,, \\
 & = & \int\frac{d^{d}p}{(2\pi)^{d}}\frac{4F(p;0)\mathcal{M}(p;0)}{p^{2}+\mathcal{M}(p;0)^{2}} \,.
\end{eqnarray}
On performing the angular integration, the corresponding expressions for
QED3 and QED4 are as follows\,:
\begin{eqnarray*}
\left\langle \bar{\psi}\psi\right\rangle _{\xi=0}^{d=3} & = & \frac{2}{\pi^{2}}\int dpp^{2}\frac{F(p;0)\mathcal{M}(p;0)}{p^{2}+\mathcal{M}(p;0)^{2}} \,,\\
\left\langle \bar{\psi}\psi\right\rangle _{\xi=0}^{d=4} & = & \frac{1}{2\pi^{2}}\int dpp^{3}\frac{F(p;0)\mathcal{M}(p;0)}{p^{2}+\mathcal{M}(p;0)^{2}} \,.
\end{eqnarray*}
For the given representative solution in the Landau gauge $\xi=0$, we find that $\left\langle \bar{\psi}\psi\right\rangle _{\xi=0}^{d=3}=(0.133\text{ GeV})^{2}$. Similarly,  $\left\langle \bar{\psi}\psi\right\rangle _{\xi=0}^{d=4}=(0.197 \text{ GeV})^{3}$.
Starting from the above solution in the Landau gauge, we apply the same procedure
as outlined in Sect. \ref{LKFEP} to calculate the fermion propagator in an arbitrary
covariant gauge both for QED3 and QED4. In case of QED3, the intermediate
position integration can be calculated analytically,~\cite{Bashir:2005wt}, leading
to the following LKF transformed fermion propagator\,:
\begin{eqnarray}
&& \hspace{-1cm}  \frac{F(p;\xi)}{p^{2}+\mathcal{M}^{2}(p;\xi)} =  \frac{a}{\pi p^{2}}\intop_{0}^{\infty}dkk^{2}\frac{F(p;0)}{k^{2}+\mathcal{M^{\mathit{2}}}(k;0)} \nonumber \\
 && \hspace{1.3cm} \times\left[\frac{1}{\lambda_{-}}+\frac{1}{\lambda_{+}}+\frac{1}{2kp}\ln\left|\frac{\lambda_{-}}{\lambda_{+}}\right|\right] \,,
\end{eqnarray}
\begin{eqnarray}
&& \hspace{-1.1cm} \frac{F(p;\xi)\mathcal{M}(p;\xi)}{p^{2}+\mathcal{M}^{2}(p;\xi)}  =  \frac{a}{\pi p}\intop_{0}^{\infty}dkk\frac{F(p;0)\mathcal{M}(p;0)}{k^{2}+\mathcal{M^{\mathit{2}}}(k;0)} \nonumber \\
 && \hspace{1.3cm} \times\left[\frac{1}{\lambda_{-}}-\frac{1}{\lambda_{+}}\right],
\end{eqnarray}
where $\lambda_{\pm}=a^{2}+(k\pm p)^{2}$ and $a=\alpha\xi/2$. We solve the above equations for several values of $\xi$ to generate $F(p;\xi)$
and $\mathcal{M}(p;\xi)$ as shown in Fig.~\ref{Fig1}. Corresponding values
of the Euclidean pole mass and fermion condensate for QED3, displayed in
Table\,1, are gauge independent within the numerical accuracy of our results. 

In the case of QED4, the position integration cannot be calculated analytically.
Consequently, we have to apply a tedious high precision numerical procedure, as the integrands are highly oscillatory, to find the fermion
propagator for different values of $\xi$. Resultant profiles of $F(p;\xi)$
and $\mathcal{M}(p;\xi)$ functions are shown in Fig. \ref{Fig2}. Corresponding
values of the Euclidean pole mass and fermion condensate given in
Table \ref{Table1}, again show their practical gauge independence. The integration formulae used in this calculation involve $p$ dependent mass function and are given by Eqs.~\ref{eq:a41} in the Appendix.

\begin{table}[h]
\begin{centering}
\begin{tabular}{|c|c|c|c|c|}
\hline 
 & \multicolumn{2}{c|}{QED3} & \multicolumn{2}{c|}{QED4}\tabularnewline
\hline 
\hline 
$\xi$ & $m_{E}$ (GeV) & $\left\langle \bar{\psi}\psi\right\rangle _{\xi}^{1/2}$ & $m_{E}$ (GeV) & $\left\langle \bar{\psi}\psi\right\rangle _{\xi}^{1/3}$\tabularnewline
\hline 
0 & 0.205 & 0.133 & 0.205 & 0.197\tabularnewline
\hline 
0.5 & 0.205 & 0.133 & 0.206 & 0.197\tabularnewline
\hline 
1.0 & 0.206 & 0.133 & 0.206 & 0.197\tabularnewline
\hline 
1.5 & 0.207 & 0.133 & 0.207 & 0.197\tabularnewline
\hline 
2.0 & 0.208 & 0.133 & 0.208 & 0.198\tabularnewline
\hline 
\end{tabular}
\par\end{centering}
\caption{Values of Euclidean pole mass and fermion condensate for different
values of the covariant gauge parameter $\xi$ in QED3 and QED4.}\label{Table1}
\end{table}

\begin{figure}[h]
\begin{centering}
\includegraphics[scale=0.9]{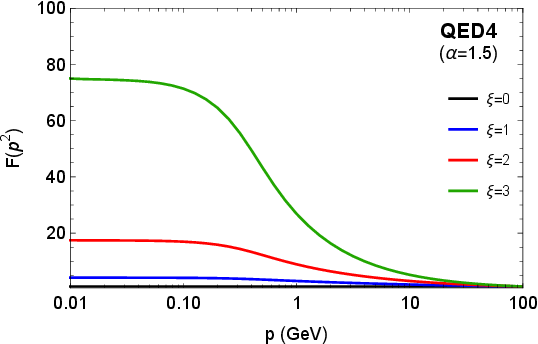}
\includegraphics[scale=0.9]{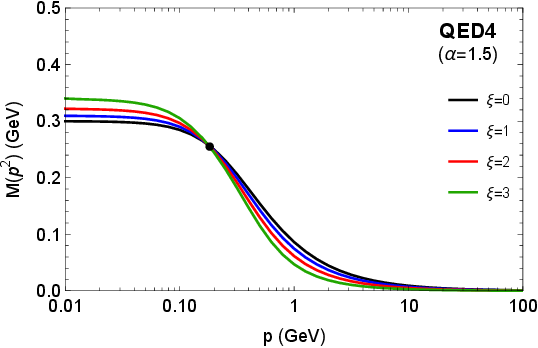}
\par\end{centering}
\caption{Wavefunction renormalization functions (above) and mass functions (below)
for various values of the gauge parameter $\xi$ for $s=1/2$ in Eq.~(\ref{model2}). The black blob in the plots
of mass functions represents the practically constant value of the Euclidean pole
mass. We take $\alpha=1.5$ and cut-off $\Lambda=10^{2}$ GeV.}\label{Fig3}
\end{figure}

 Strict gauge-independence of the fermion condensate in QED was firmly established in~\cite{Bashir:2005wt}. What about the fermion pole mass? Its gauge-invariance  was explicitly demonstrated in~\cite{DallOlio:2019cpz}.
This proof was based upon the Nielsen-idenities~\cite{Nielsen:1975fs,Breckenridge:1994gs}. One might consider LKF transformations as the integrated version of the Nielsen identities~\cite{DallOlio:2019cpz}. Therefore, the apparent gauge-invariance of the pole mass in Fig.~\ref{Fig1} and Fig.~\ref{Fig2} is not a surprise. In order to verify this, we generalize our model for the mass function along the suggestions made in~\cite{Bashir:2011ij}:
\begin{eqnarray}
\mathcal{M}(p;0) = m \left( \frac{m^2}{m^2+p^2}\right)^{s} \,. \label{model2}
\end{eqnarray}
The model we have already investigated corresponded to the choice $s=1$. However, near the critical coupling in quenched QED, the anomalous mass dimension $\gamma_m$, defined as, 
\begin{eqnarray}
M(p;0) \propto (p^2)^{{\gamma_m/2}-1} \,.
\end{eqnarray}
in the deep Euclidean region ($p^2 \rightarrow \infty $) has been argued to be $\gamma_m \sim 1$,~\cite{Holdom:1988gs}. This corresponds to $s=1/2$ in our generalized model. We repeat the LKF transformation with this choice of $s$ and find that the pole mass continues to be gauge-independent as depicted in Fig.~\ref{Fig3}. 

\section{Conclusions}\label{conclusion}

We provide a detailed analysis and application of the LKF transformation in QED both for massless and massive fermion propagator in three and four space-time dimensions. Though several expressions have been derived in the literature before, we encompass all the cases and extend our results to two loops, including the renormalized results in the $\overline{\text{MS}}$ scheme. We compare our renormalization constants with earlier efforts in the field~\cite{DallOlio:2021njq} and find agreement. Though this latter work, which in turn agrees with the direct two loop computation of the renormailzation constants reported in~\cite{Fleischer:1998dw}, is for the quark propagator in QCD, setting the Casimir color factors $C_F=1$ and $C_A=0$ provides the results for QED.   
We also provide LKF transformed results for a representative functional form of the dynamically generated mass function in quenched QED3 and QED4. Though the fermion propagator does vary with the variation of gauge, as expected, the chiral fermion-anti-fermion condensate is strictly gauge-independent although the mass function varies slightly with the variation of the covariant 
gauge parameter. Our work sets the stage for a complete numerical analysis of the gauge transformation of the dynamically generated mass function in QED with sophisticated truncations available, see for example~\cite{Bashir:2011dp,Albino:2018ncl}. It will invoke the the implementation of transverse Takahashi identities as well in order to ensure several implications of gauge invariance are taken into account.

The eventual final step is to carry out the task in QCD~\cite{Aslam:2015nia,Bermudez:2017bpx,Sultan:2018tet,DeMeerleer:2018txc,DeMeerleer:2019kmh,DallOlio:2021njq,Lessa:2022wqc}, and study the gauge variation of the quark propagator, the quark mass function, its wavefunction renormalization and establish the gauge independence of the associated hadronic observables. All this is for future.

\section*{Acknowledgements}

A.~B. wishes to acknowledge the {\em Coordinaci\'on de la Investigaci\'on Cient\'ifica} of the{\em Universidad Michoacana de San Nicol\'as de Hidalgo} grant 4.10, the
{\em Consejo Nacional de Humanidades, Ciencias y Tecnolog\'ias} project
CBF2023-2024-3544
as well as the Beatriz-Galindo support during his current scentific stay at the University of Huelva, Huelva, Spain. F.~A. acknowledges HEC grant 20-15728/NRPU/R\&D/HEC/2021, Pakistan.

\appendix
\numberwithin{equation}{section}
\section{Useful integrals and identities}\label{AppendixA}
In this Appendix, we collect useful integrals, identities and properties of confluent Hyper-geometric functions used in the main text of the article.
Note that the angular integrals involved in this work are of the following form\,:
\begin{equation}
\int_{0}^{\pi} \text{e}^{-ipx\cos\theta}\cos\theta\sin\theta d\theta = 2i\left(\frac{\cos px}{px}-\frac{\sin px}{p^2x^2}\right)\,,\label{eq:A1}
\end{equation}
while the radial integration in momentum space invokes the following standard identities\,:
\begin{eqnarray}
    \int_{0}^{\infty} p^{n} \cos px\;dp &=& -(x)^{-(n+1)}\Gamma\left(1+n\right)\sin\frac{n\pi}{2}\, ,\notag\\
     \int_{0}^{\infty} p^{n} \sin px\;dp &=& (x)^{-(n+1)}\Gamma\left(1+n\right)\cos\frac{n\pi}{2}\, . \quad
     \label{eq:A2}
\end{eqnarray}
For the massive case in three dimensions, the following radial integrals are made use of\,: 
\begin{eqnarray}
 \int_{0}^{\infty} \frac{p^2}{p^2+m^2} \cos px\;dp &=& -\frac{m\pi\text{e}^{-mx}}{2}\,,\notag\\
 \int_{0}^{\infty} \frac{p}{p^2+m^2} \sin px\;dp &=& \frac{\pi\text{e}^{-mx}}{2}\,.\label{eq:A3}
\end{eqnarray}
To perform the radial integration in Eq.~(\ref{eq:23}), we use
\begin{eqnarray}
 \int dx\;\text{e}^{-mx}\sin px&=&\frac{p}{p^{2}+m^{2}}\,, \notag\\
 \int dx\;x\text{e}^{-mx}\sin px&=&\frac{2mp}{\left(p^{2}+m^{2}\right)^{2}}\,,\notag\\
 \int dx\;x^{2}\text{e}^{-mx}\sin px&=&-\frac{2p\left(p^{2}-3m^{2}\right)}{\left(p^{2}+m^{2}\right)^{3}}\,,\label{eq:A3}
\end{eqnarray}
whereas for $B\left(p;\xi\right)$, expressed through  Eq.~(\ref{eq:24}), we employ the following standard formulae\,:
\begin{eqnarray}
 \int dx\;\text{e}^{-mx}\cos px&=&\frac{m}{p^{2}+m^{2}}\,, \notag\\
 \int dx\;x\text{e}^{-mx}\cos px &=&-\frac{p^2-m^2}{\left(p^{2}+m^{2}\right)^{2}}\,,\notag\\
 \int dx\;x^{2}\text{e}^{-mx}\cos px &=&-\frac{2m\left(3p^{2}-m^{2}\right)}{\left(p^{2}+m^{2}\right)^{3}}\,, \notag \\
  \int dx\;x^{3}\text{e}^{-mx}\cos px &=&-\frac{6\left(m^4-6m^2p^{2}+p^{4}\right)}{\left(p^{2}+m^{2}\right)^{4}}\,. \label{eq:A4}
\qquad 
\end{eqnarray}
The Hyper-geometric functions can be expanded in the powers of $p^2/m^2$, and at the leading order, one gets \cite{Bashir:2002sp}:
\begin{align}
_{2}F_{1}\left(2-\nu,1-\nu;2;-\frac{p^{2}}{m^{2}}\right) & =1-\frac{\left(2-\nu\right)\left(1-\nu\right)}{2}\frac{p^{2}}{m^{2}}\,,\nonumber \\
_{2}F_{1}^{2}\left(2-\nu,1-\nu;2;-\frac{p^{2}}{m^{2}}\right) & =1-\left(2-\nu\right)\left(1-\nu\right)\frac{p^{2}}{m^{2}}\,,\nonumber \\
_{2}F_{1}\left(3-\nu,1-\nu;3;-\frac{p^{2}}{m^{2}}\right) & =1-\frac{\left(3-\nu\right)\left(1-\nu\right)}{3}\frac{p^{2}}{m^{2}}\,,\nonumber \\
_{2}F_{1}^{2}\left(3-\nu,1-\nu;3;-\frac{p^{2}}{m^{2}}\right) & =1-2\frac{\left(3-\nu\right)\left(1-\nu\right)}{3}\frac{p^{2}}{m^{2}}\,.\label{eq:A5}
\end{align}
For $n\geq2$, we can write
\begin{eqnarray}
&& \hspace{-1.2cm}  \Gamma\left(n+\varepsilon\right) =\Gamma\left(1+\varepsilon\right)\Gamma\left(n\right)   \notag \\
&&  \hspace{-0.5cm} \times \left[1+\varepsilon \mathcal{Z}_{1}\left(n-1\right)+\cdots+\varepsilon^{n-1}\mathcal{Z}_{1\ldots1}\left(n-1\right)\right]\,.
\label{eq:A10}
\end{eqnarray}
With this general relation, we can check:
\begin{eqnarray}
\Gamma\left(2-\nu\right) & =&\Gamma\left(1-\nu\right)\left[1-\nu \mathcal{Z}_{1}\left(1\right)\right]\,,\label{eq:A11}\\
\Gamma\left(3-\nu\right) & =&\left(1-\nu\right)\Gamma\left(3\right)\left[1-\nu \mathcal{Z}_{1}\left(2\right)+\nu^{2}\mathcal{Z}_{11}\left(2\right)\right]\,.\notag
\end{eqnarray}
Using
\begin{equation*}
\vspace{-2mm}
\mathcal{Z}_{1}\left(n-1\right)=\sum_{i=1}^{n-1}\frac{1}{i}\,,
\end{equation*}
we can show that
\begin{align}
\mathcal{Z}_{1}\left(1\right) & =\frac{1}{1}=1;\quad \mathcal{Z}_{1}\left(2\right)=\frac{1}{1}+\frac{1}{2}=\frac{3}{2}\,,\nonumber \\
\mathcal{Z}_{11}\left(2\right) & =\sum_{2\geq i_{1}>i_{1}>0}\frac{1}{i_{1}}\frac{1}{i_{2}}=\frac{1}{2}\frac{1}{1}=\frac{1}{2}\,.\label{eq:40}
\end{align}
It allows us to write Eq.~(\ref{eq:A11})\,:
\begin{align}
\Gamma\left(2-\nu\right) & =\Gamma\left(1-\nu\right)\left[1-\nu\right]\,,\nonumber \\
\Gamma\left(3-\nu\right) & =\Gamma\left(1-\nu\right)2\left[1-\frac{3}{2}\nu+\frac{1}{2}\nu^{2}\right]\,.\label{eq:41}
\end{align}
The integral formula of four transformed and the inverse four transformed for variable mass function in
case of QED4 are as follow :
\begin{align}
X(x,0) & =  -\frac{1}{4\pi^{2}x^{2}}\intop_{0}^{\infty}dpp^{3}\frac{F(p;0)}{p^{2}+\mathcal{M}^{2}(p;0)}J_{2}(px)\,,\nonumber \\
Y(x,0)  & =  -\frac{1}{4\pi x}\intop_{0}^{\infty}dpp^{2}\frac{F(p;0)\mathcal{M}(p;0)}{p^{2}+\mathcal{M}^{2}(p;0)}J_{1}(px)\,,\nonumber \\
A(p;\xi) & =  \frac{4\pi^{2}}{p}\intop_{0}^{\infty}dxx^{2}Y(x;\xi)J_{1}(px)\,, \nonumber \\
B(p;\xi) & =  4\pi^{2}\intop_{0}^{\infty}dxx^{3}X(x;\xi)J_{2}(px) \,.\label{eq:a41}
\end{align}

\bibliography{LKF-QED2Loop}
\bibliographystyle{andr-jam.bst}

\end{document}